\documentclass{PoS}

\usepackage{graphicx}
\usepackage{amsmath}
\usepackage{amssymb}
\usepackage{multirow}

\newcommand{\smc}[1]{ {\scriptstyle{\text{#1}}} }

\DeclareMathOperator{\Tr}{Tr}

\title{Predictions for LHC from SO(4) MWT}

\ShortTitle{Predictions for LHC from SO(4) MWT}

\author{\speaker{Ari Hietanen}\\
        E-mail: \email{hietanen@cp3-origins.net}}

\author{Claudio Pica\\
        E-mail: \email{pica@cp3-origins.net}}

\author{Francesco Sannino\\
        E-mail: \email{sannino@cp3-origins.net}}

\author{Ulrik S{\o}ndergaard\\
        E-mail: \email{sondergaard@cp3-origins.net}}

\author{\\CP$^3$-Origins \& the Danish Institute for Advanced Study DIAS,
        University of Southern Denmark, Campusvej 55, DK-5230 Odense M, Denmark.}

\abstract{We investigate the vector, axial and pseudo scalar mass
  spectrum of an SO(4) - MWT gauge theory with fermions in the vector
  representation of SO(4).  Here we present the preliminary  lattice
  results for the masses of vector and axial vector meson using Wilson
  fermions. These spectra are crucial for the discovery and to guide
  the searches of composite dynamics at the LHC.\\\\
  CP3-Origins-2013-045 DNRF90 \& DIAS-2013-45
} 

\FullConference{31st International Symposium on Lattice Field Theory - LATTICE 2013\\
		July 29 - August 3, 2013\\
		Mainz, Germany}

\begin{document}

\section{Introduction}
In Technicolor (TC) theories a new a strongly interacting gauge theory
is responsible for the electroweak symmetry breaking phenomenon. This scenario
leads to a spectrum of composite particles with masses in the TeV energy range. The specific spectral properties of 
the composite states can be calculated via lattice simulations and depend on the specific underlying gauge theory. The first general classification of SU($N$) gauge theories that can lead to phenomenologically acceptable composite electroweak theories appeared in \cite{Sannino:2004qp,Dietrich:2006cm,Pica:2010xq}, including the presence of a light composite Higgs-like state \cite{Sannino:2003xe,Hong:2004td,Dietrich:2005jn}. The generalization to SO, Sp  and exceptional groups appeared in \cite{Sannino:2009aw,Mojaza:2012zd}.

Furthermore it has been recently argued that a phenomenologically viable TC Higgs mass can arise via a  cancellation between the intrinsic positive mass squared of the composite state and the large negative top corrections \cite{Foadi:2012bb}. It is relevant to test these expectations via lattice simulations. However the isosinglet scalar meson is challenging to measure on lattice because of noisy disconnected diagrams, and furthermore one should add four-fermion interactions allowing the generation of the top mass operator. It is for these reasons that we turn our attention first to the spectrum of isotriplet spin-one resonances whose spectral properties can be determined more reliably via lattice simulations, and furthermore,  their masses are expected to  remain close to the one directly determined via lattice simulations, in units of the electroweak scale (i.e. pion decay constant). 
%

In this proceedings we provide preliminary results for the
physical values of vector and axial vector meson from the SO(4) vector
Minimal Walking Theory (MWT) at a fixed lattice spacing. Earlier data are already
published in \cite{Hietanen:2012sz,Hietanen:2012qd}.  The SO(4) vector MWT is phenomenologically interesting theory for several
reasons. First, from perturbative arguments the theory is expected to
break chiral symmetry, but to be near the lower bound of
conformal window  \cite{Sannino:2009aw}. Second, the theory contains possibly light composite
dark matter particles. These new light states emerge because the techni-fermions belong to a real
representation, and therefore the chiral symmetry breaking pattern is expected to be SU(4)$\rightarrow$SO(4). This leads to nine Goldstone bosons. Three of
these are eaten by the SM gauge bosons leaving six additional Goldstone bosons forming an
electroweak complex triplet. The
neutral isospin zero component is a possible dark matter candidate
called ITIMP \cite{Frandsen:2009mi}. Third, compared to the SU(2)$_{\rm
  adj}$-MWT using the standard hyper charge assignments there are no
fractionally charged composite particles as technigluons cannot form bound
states with a techniquark. The spectrum of the TC theories have also been studied in
\cite{Fodor:2011tu,Fodor:2012ty} and theories with composite dark 
matter in \cite{Lewis:2011zb,Appelquist:2013ms,Hietanen:2013fya}.  
%

\section{SO(4) gauge theory with two fundamental on the Lattice}
 We used, in our simulations, the Wilson plaquette action with
unimproved Wilson fermions
\begin{equation}
S=S_{\rm F} +S_{\rm G}, ,
\end{equation}
where
\begin{equation}
S_{\rm G}= \beta \sum_x \sum_{\mu,\nu < \mu} \left[  1- \frac{1}{N_{\rm
    c}} \Tr U_{\mu\nu}(x) \right] \, , \label{eq:Sg}
\end{equation}
is the Yang-Mills gauge action.  $U_{\mu\nu}(x)$ is the plaquette
defined in terms of the link variables as 
\begin{equation}
U_{\mu\nu}(x)=U_\mu(x)U_\nu(x+\hat \mu)U^T_\mu(x+\hat \mu+\hat \nu)U^T_\nu(x+\hat \nu) \, .
\end{equation}
The Wilson fermion action is \begin{equation}
S_\smc{F}=\sum_{f} \sum_{x,y}  \bar \psi_f(x) M(x,y) \psi_f(y) \ ,
\end{equation}
with $f$ running over fermion flavors and the Wilson-Dirac matrix $M(x,y)$ given by
\begin{align}
\begin{split}
\sum_{y}M&(x,y) \psi(y) = (4+m_0)\psi (x)
- \frac{1}{2}\sum_{\mu}\Big[(1+\gamma_\mu)U^T_\mu(x-\hat\mu)\psi(x-\hat\mu)  +   (1-\gamma_\mu)U_\mu(x)\psi(x+\hat\mu) \Big] \, .
\end{split} \label{eq:Wilson-Dirac}
\end{align}
The observables we are interested in are the PCAC mass, pseudo scalar meson,
vector meson, and axial vector meson mass as well as the pseudo scalar
decay constant defined in \cite{Hietanen:2012sz}. 

\section{Lattice Results}
\begin{figure}
  \begin{center}
    \includegraphics[width=0.65\textwidth]{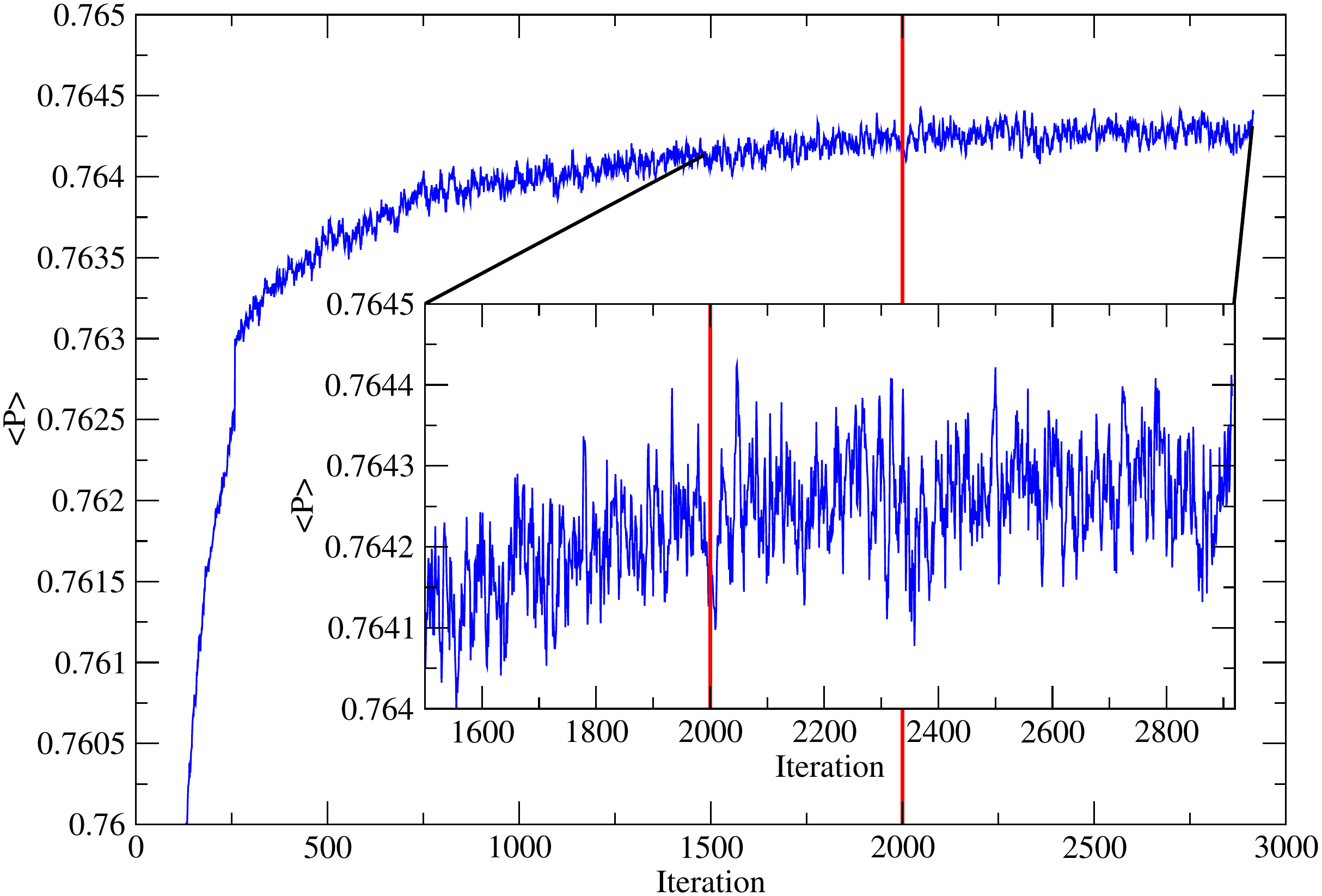}
    \caption{Thermalization of Plaquette expectation values when
      starting from a random configurations. The first 2000
      configurations have been discarded in the analysis.\label{fig:plaq_therm}} 
  \end{center}
\end{figure}

\begin{figure}
  \begin{center}
    \includegraphics[width=0.45\textwidth]{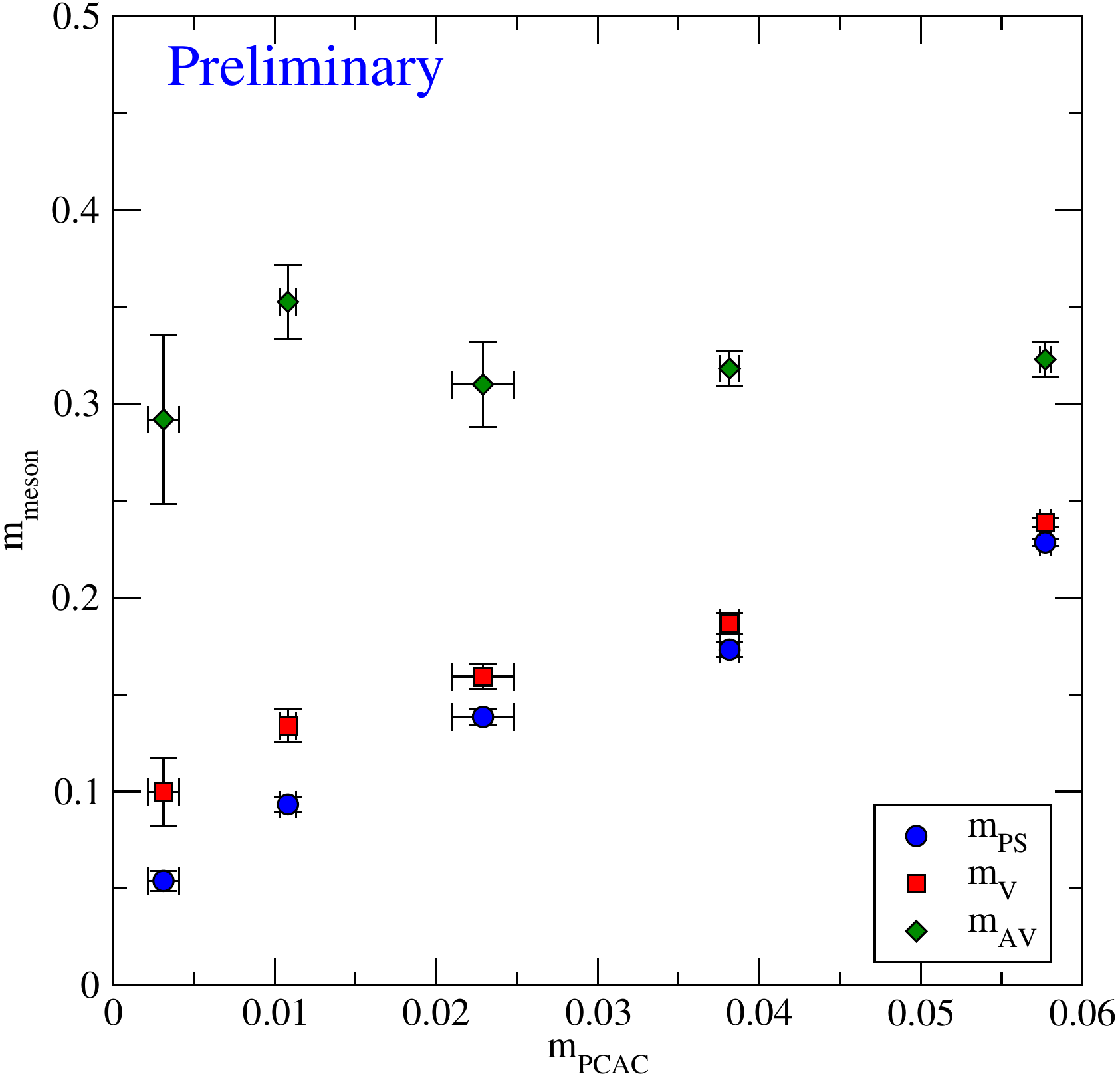}
    \caption{The masses of pseudo scalar, vector, and axial vector
      meson as a function of PCAC-quark mass.\label{fig:spectrum}} 
  \end{center}
\end{figure}

\begin{figure}
  \begin{center}
    \includegraphics[width=0.45\textwidth]{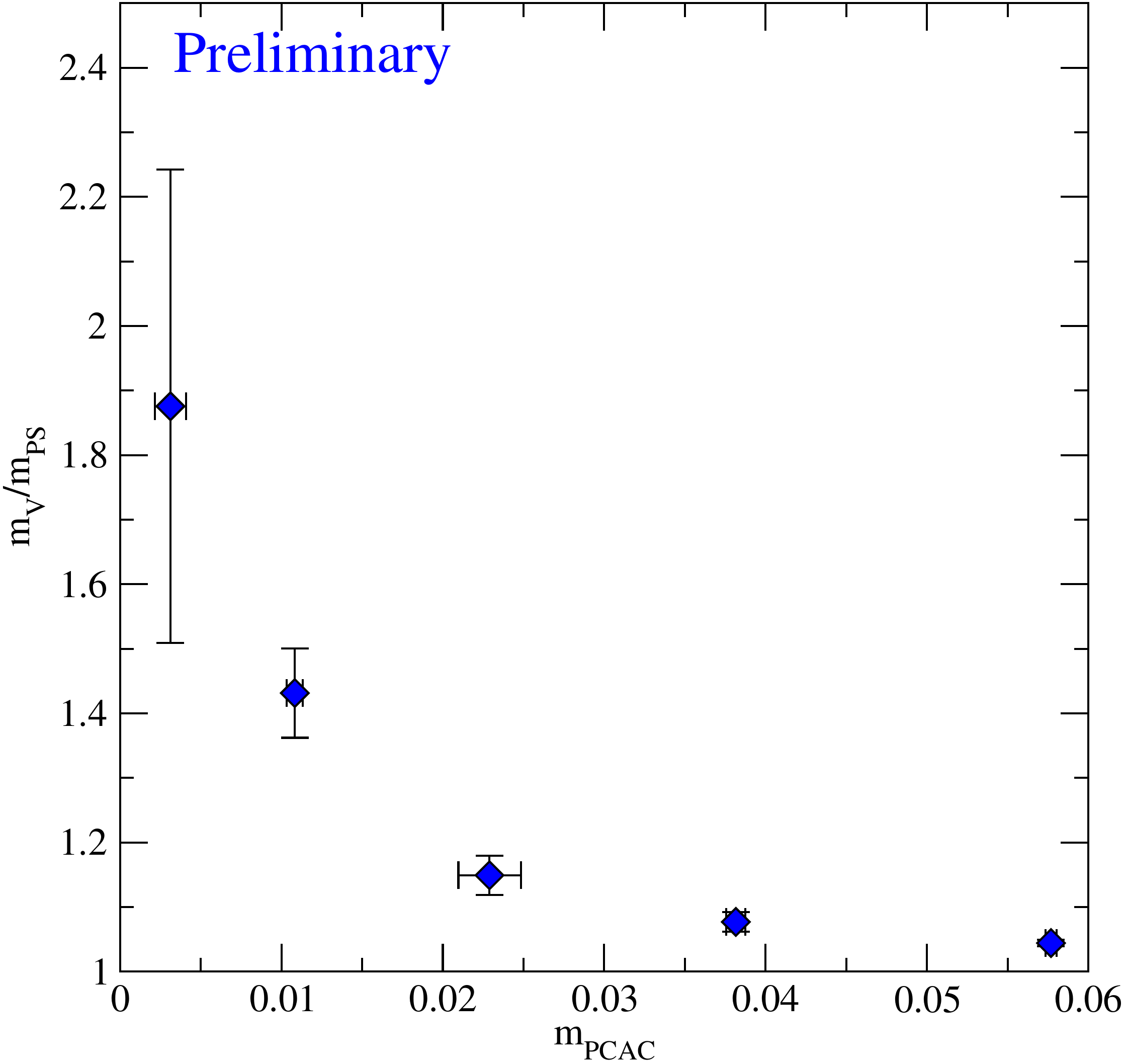}
    \includegraphics[width=0.45\textwidth]{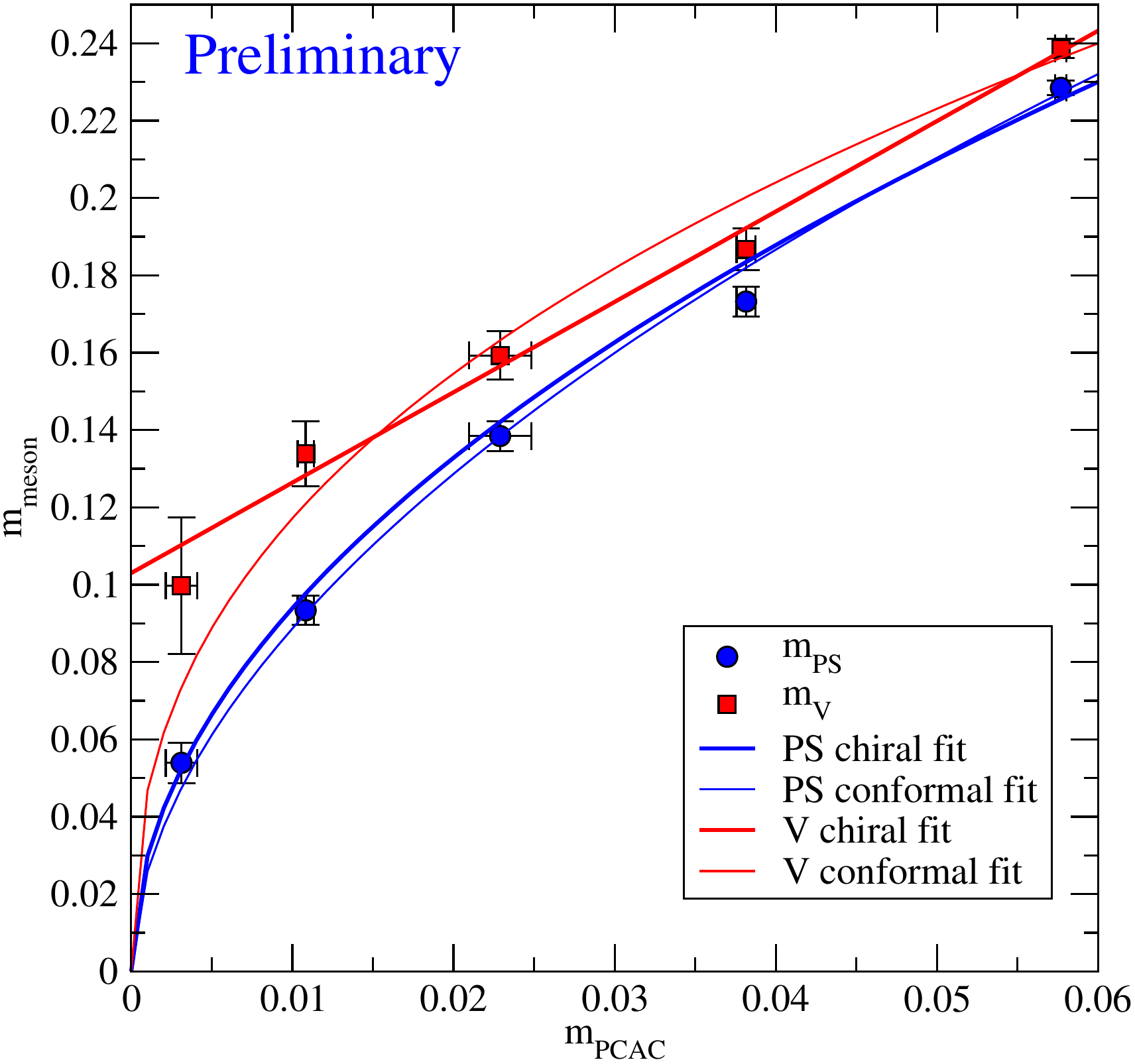}
    \caption{Left: Ratio between vector  and pseudo scalar
      meson. Right: Chiral (thick lines) and conformal (thin lines)
      fits to the pseudo scalar and vector meson masses. \label{fig:ratio}}
  \end{center}
\end{figure}

\begin{figure}
  \begin{center}
    \includegraphics[width=0.45\textwidth]{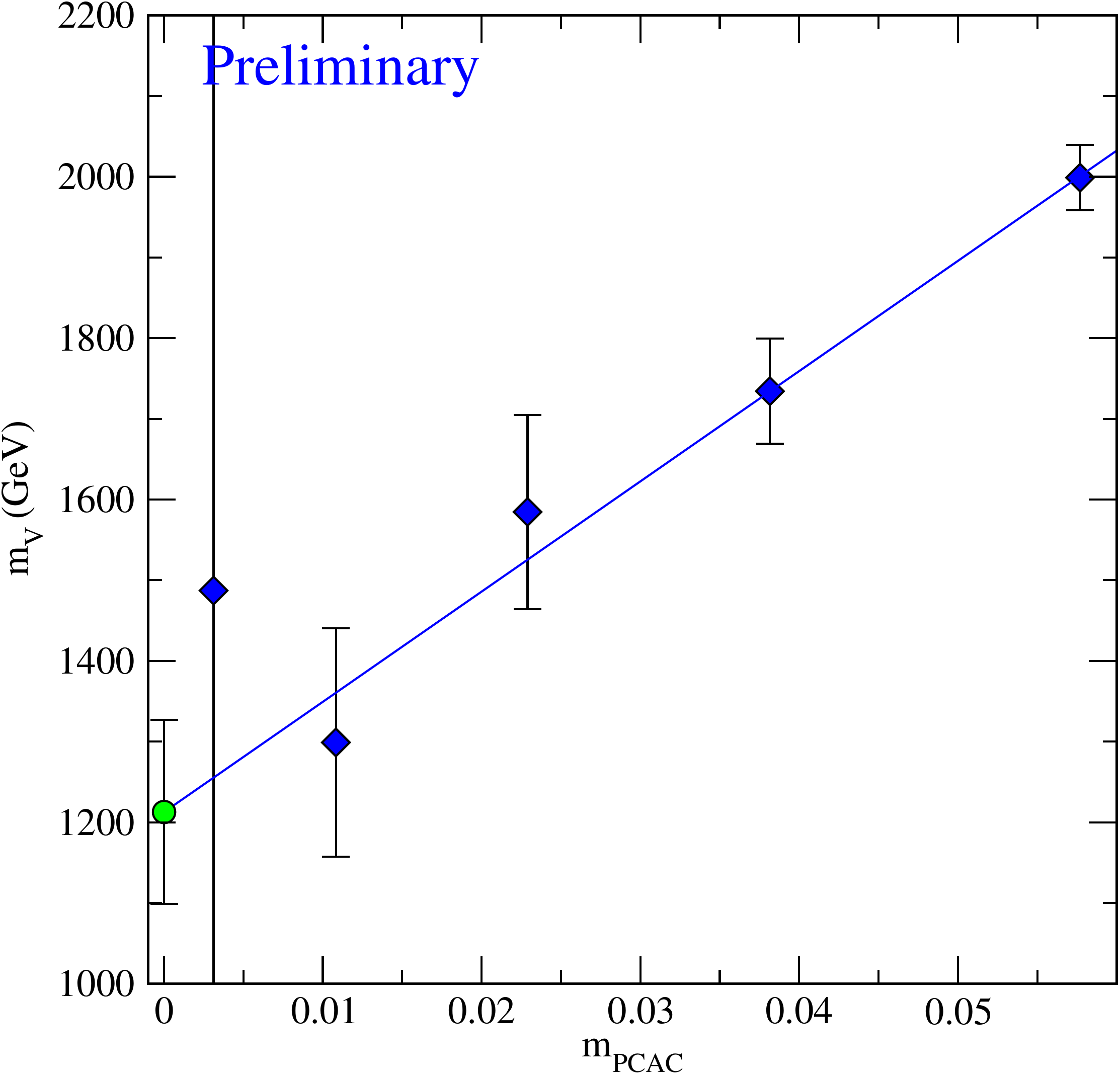}
    \caption{Chiral extrapolation of vector meson mass in physical
      units: $m_v = m_{\rm V}^{\rm lat}/f^{lat}_{\rm PS} 246$
      GeV. \label{fig:spectrum_phys}}
  \end{center}
\end{figure}

In this proceedings we present the preliminary results for one
lattice spacing $\beta=7$ in the chiral region for
$V=64\times32^3$. Compared to the results at smaller volumes
$V=64\times24^3$ \cite{Hietanen:2012sz} we observe considerable finite
volume effects. However, the conclusion suggesting the occurrence of chiral symmetry
breaking is unaffected.  

The configurations for the heavier masses were generated using
standard HMC algorithm. For the two points closest to the chiral limit
Hasenbush preconditioning was used \cite{Hasenbusch:2001ne}. The
trajectory length was chosen to be one.  The thermalization of the
simulations is slow and takes about 2000 iterations. An example of
thermalization of Plaquette is shown in Fig.~\ref{fig:plaq_therm} for
$m_0=-0.3$. Other quantities have similar long tails, 
approaching the correct value slowly. The auto correlation time is
about 10 iterations.

In Fig.~\ref{fig:spectrum} we plot the masses of the pseudo
scalar, vector, and axial vector mesons. The three lightest data points have
low statistics with the shown statistical errorbars probably underestimates. However, there is a clear separation between 
different particle states indicating the occurrence of chiral symmetry breaking. This
can also be seen from the left panel of Fig.~\ref{fig:ratio} where we plot the ratio of vector to pseudo scalar meson mass, which
diverges in the chiral limit. 

For a more systematic comparison between the possible occurrence of chiral symmetry and the hypothesis of large distance conformality, we perform also a {\it conformal} fit
\begin{equation}
  m_{\rm PS, V,AV} = A m_{\rm q}^{\frac{1}{1+\gamma}}\, ,
\end{equation}
where $0\le\gamma\le2$ is a universal exponent.  The chiral symmetry breaking scenario would, on the other hand,  predict the following relations among the measured hadron masses as function of the quark mass:
\begin{align}
  \frac{m^2_{\rm PS}}{m_{\rm q}} & = A + \mathcal{O}\left(m_{\rm q}
  \log(m_{\rm q})\right)\nonumber\\
  m_{\rm V,AV} & = A + B m_{\rm q} + \mathcal{O}\left(m_{\rm q}
  \log(m_{\rm q})\right).
\end{align}
The different parameters and $\chi^2/$dof for the fits are given in
the Table~\ref{tbl:fits}.  The scenario of chiral symmetry
breaking seems favored. The axial vector meson dependence on the underlying fermion mass is, for example, clearly incompatible with the conformal fit. The  different fits for pseudo scalar and vector
meson masses are  represented in the right panel of Fig.~\ref{fig:ratio}.

\begin{table}
\begin{center}
\begin{tabular}{llll}
  \hline
  Meson & Fit function\;\;\; & Best parameter\;\;\; & $\chi^2/$dof   \\
  \hline
  \hline
  PS ChSB &   $A\sqrt{m}$ & $A=0.939(7)$ & 11.7/4 \\
  \multirow{2}{*}{PS conformal} & \multirow{2}{*}{$Am^\frac{1}{1+\gamma}$} & $A=1.05(6)$  &  \multirow{2}{*}{7.2/3} \\
  &   & $\gamma=0.86(4)$ &\\
  \hline
  \multirow{2}{*}{Vector ChSB} &  \multirow{2}{*}{$A + Bm$} & $A=0.103(7)$ &  \multirow{2}{*}{2.1/3}  \\
  & & $B=2.34(12)$ &\\
  \multirow{2}{*}{Vector conformal\;\;\;} & \multirow{2}{*}{$Am^\frac{1}{1+\gamma}$} & $\gamma=1.50(7)$  &  12.1/3 \\
  & & $A=0.74(6)$ & \\
  \hline
  \multirow{2}{*}{Axial vector ChSB} &  \multirow{2}{*}{$A + Bm$} & $A=0.103(7)$ &  \multirow{2}{*}{2.1/3}  \\
  & & $B=2.34(12)$ &\\
  \multirow{2}{*}{Axial Vector conformal\;\;\;} & \multirow{2}{*}{$Am^\frac{1}{1+\gamma}$} & $\gamma=2$  &  93.3/3 \\
  & & $A=0.924(17)$ & \\
  \hline
\end{tabular}
\caption{Different types of fit functions in the chiral regime for the data with $m$ identified with the $m_\smc{PCAC}$.\label{tbl:fits}}
\end{center}
\end{table}

To determine the physical values of the masses we set the pseudo scalar decay
constant to $246$~GeV to recover the physical weak gauge boson masses. After performing
a chiral extrapolation with these values we obtain a mass of $m_V =
1.21(11)$TeV, as shown in Fig.~\ref{fig:spectrum_phys}. Similarly for the axial mass we have $m_A=3.5(3)TeV$. Potentially large unknown corrections from the continuum extrapolation could arise requiring, in the future, to provide new measurements at another value of $\beta$. In addition, we  still need the renormalization constant $Z_a$ which enters in the denominator of the pseudo
scalar decay constant. The value from perturbation theory for $Z_a$
in SO($N$) gauge theories with vector representation fermions is given in \cite{DelDebbio:2008wb}
\begin{equation}
Z_a=1-\frac{g_0^2}{16\pi^2}\frac{N-1}{2}15.7\overset{N=4}{=}1-1.1931/\beta
\overset{\beta=7}{=} 0.823.
\end{equation}
The perturbative correction would increase the physical vector meson
mass to 1.47(14)TeV.

\section{Summary}
We have presented preliminary results of the simulations of  the SO($4$)-MWT with $\beta=7$ and $V=64\times32^3$. The data seem to strongly suggest the occurrence of 
chiral symmetry breaking. In addition, we have also obtained a
preliminary predictions for the vector and axial vector meson
masses. The vector meson mass 1.47(14)TeV  is quite close to that one
of  the $\rho$ meson in QCD obtained by setting $f_{\rm \pi}=246$GeV where
as the axial vector mass is much heavier. Such a vector meson
mass should be within LHC experimental reach. However, although intriguing, these results need further confirmation by going beyond the one lattice spacing approximation. At 
the moment we are improving the results by performing more simulations
near the chiral limit as well as with an another lattice spacing.   

\subsection*{Acknowlegements}
This work was supported by the Danish National Research Foundation DNRF:90
grant and by a Lundbeck Foundation Fellowship grant. We also
acknowledge PRACE for awarding us access to resource 
on MARENOSTRUM based in Spain at BSC. Additional computing facilities
were provided by the Danish Centre for Scientific Computing.

\end{document}